\documentclass{PoS}

\usepackage{cite}

\title{On the ratio of $t\bar{t}\gamma$ and $t\bar{t}$ cross sections at the LHC}

\ShortTitle{On the ratio of $t\bar{t}\gamma$ and $t\bar{t}$ cross sections at the LHC}

\author{
\speaker{Giuseppe Bevilacqua}\thanks{Work supported by grant K 125105 of the National Research, Development and Innovation Office in Hungary.} \\
        MTA-DE Particle Physics Research Group, H-4002 Debrecen, PO Box 400, Hungary \\
        E-mail: \email{giuseppe.bevilacqua@science.unideb.hu} 
}

\abstract{We study the ratio of the cross sections for $t\bar{t}\gamma$ and $t\bar{t}$ production at the LHC. The presence of correlations between theoretical uncertainties of the two processes makes possible a precise determination of this observable. This can help to evidentiate effects of new physics that might reveal themselves only when sufficiently precise theoretical predictions are available. Our analysis is based on fully realistic simulations of $t\bar{t}\gamma$ and $t\bar{t}$ production in the dilepton decay channel, including complete off-shell and non-resonant effects at NLO QCD accuracy. We discuss Standard Model predictions for the LHC Run II at both inclusive and differential level, also quantifying the impact of the theoretical uncertainties related to variation of scales and parton distribution functions.}

\FullConference{XXVII International Workshop on Deep-Inelastic Scattering and Related Subjects - DIS2019 \\
		8-12 April, 2019\\
		Torino, Italy}

\begin{document}

\section{Introduction}

Precise measurements of top quark properties provide important tests of the Standard Model (SM) as well as opportunities to probe effects of new physics. As top quark pairs are produced abundantly at the LHC, the cross section of the process $pp \to t\bar{t} + X$ is a natural benchmark for these studies. At present, the inclusive $t\bar{t}$ production process is known with accuracy NNLO QCD + NNLL$^\prime$ combined with electroweak corrections \cite{Czakon:2013goa, Czakon:2017wor}. On a more exclusive ground, the first NNLO QCD results for $t\bar{t}$ production with top quark decays in the narrow width approximation have recently started to appear \cite{Behring:2019iiv}.
With the large amount of data collected during the Run II of the LHC, rarer production channels such as  $t\bar{t}H$ or $t\bar{t}V$ ($V= \gamma, W^\pm, Z$), become also accessible \cite{Sirunyan:2018hoz,Sirunyan:2017uzs,Aaboud:2018hip}. Among these, $t\bar{t}\gamma$ production is most sensitive to the top quark electromagnetic coupling. Deviations with respect to SM predictions in the $p_T$ spectrum of the photon could unveil new physics through anomalous top quark couplings \cite{Baur:2004uw,Bylund:2016phk,Schulze:2016qas}. Clearly, precise theoretical predictions are a crucial ingredient for such measurements. 

The present state of the art for the theoretical description of $t\bar{t}\gamma$ is NLO.
Both QCD and EW corrections are available for fully inclusive on-shell production \cite{PengFei:2009ph,PengFei:2011qg,Duan:2016qlc,Maltoni:2015ena}. Exclusive predictions have been calculated in the narrow width approximation, with NLO QCD  corrections applied to both production and decay stages \cite{Melnikov:2011ta}. Results for on-shell $t\bar{t}\gamma$ production matched to parton showers are also available \cite{Kardos:2014zba}. More recently, NLO QCD predictions to the full process $pp \to e^+ \nu_e \mu^- \bar{\nu}_\mu b \bar{b} \, \gamma + X$ have been computed \cite{Bevilacqua:2018woc}. The latter provide the most realistic description of $t\bar{t}\gamma$ production and decay in the dilepton channel at fixed perturbative order, including complete off-shell effects and non-resonant contributions.

Going beyond NLO is extremely challeging for $t\bar{t}\gamma$. Achieving a full NNLO calculation could reduce theoretical uncertainties down to a few percents, paving the road to a precision study of this process. Though highly desirable, this task is beyond reach at present as $2 \to 3$ processes represent the frontier of NNLO. \textit{A fortiori}, applications of NNLO to processes with higher particle multiplicity like $W^+W^-b\bar{b}\gamma$ (\textit{i.e.} off-shell $t\bar{t}\gamma$) are  difficult to imagine in the near future. One class of observables which may feature reduced uncertainties using state-of-the-art theoretical accuracy is represented by \textit{cross section ratios}. Ratios may be significantly more stable against radiative corrections and scale/PDF variations than absolute quantities \cite{Melnikov:2011ta,Mangano:2012mh,Bevilacqua:2014qfa,Schulze:2016qas}. 
Given its large cross section and similar behaviour with respect to radiative corrections \cite{Denner:2010jp,Bevilacqua:2010qb,Denner:2012yc}, $t\bar{t}$ production is a particularly suitable candidate for normalizing the absolute $t\bar{t}\gamma$ predictions. In this contribution we present selected results from our recent work \cite{Bevilacqua:2018dny} where we focus on the observable
$
\mathcal{R} = \sigma_{t\bar{t}\gamma}/\sigma_{t\bar{t}}
$
and explore its performance in accuracy in the context of the LHC Run II.

\section{Details of the calculation}

Our analysis is based on a full NLO QCD calculation of the processes $pp \to e^+ \nu_e \mu^- \bar{\nu}_\mu b \bar{b} \, \gamma + X$ and $pp \to e^+ \nu_e \mu^- \bar{\nu}_\mu b \bar{b} + X$ at the center-of-mass evergy of 13 TeV.
In each case, all resonant and non-resonant Feynman diagrams, interferences and finite-width effects of the top quark and $W$ boson decays have been taken into account. For ease of notation, in the following we will also refer to these two processes as to "$t\bar{t}\gamma$" and "$t\bar{t}$" respectively. For the sake of clarity let us stress that with "absolute $t\bar{t}\gamma$ predictions" and "$t\bar{t}\gamma/t\bar{t}$ ratio"  we mean quantities calculated out of realistic final states with $b$-jets, charged leptons and missing $p_T$, and that no on-shell approximation is applied to the intermediate top or $W$ resonances.
Details about the SM parameters used for the calculation are described in Ref. \cite{Bevilacqua:2018dny}.
Following the PDF4LHC recommendations for LHC Run II \cite{Butterworth:2015oua}, we consider the PDF sets CT14 \cite{Dulat:2015mca}, MMHT14 \cite{Harland-Lang:2014zoa} and NNPDF3.0 \cite{Ball:2014uwa} for our predictions.

On the technical side, our results have been obtained with the help of the package \texttt{HELAC-NLO} \cite{Bevilacqua:2011xh}, which comprises \texttt{HELAC-1LOOP} \cite{vanHameren:2009dr} and \texttt{HELAC-DIPOLES} \cite{Czakon:2009ss,Bevilacqua:2013iha}. 
The final results are available in the form of events in either Les Houches Event File format or ROOT Ntuples that might be directly used for experimental studies at the LHC. Each event is stored with additional matrix-element and PDF information to allow reweighting for different scale and PDF choices \cite{Bern:2013zja}. A newly developed tool, called \texttt{HEPlot}, can be used to get predictions from these Ntuples for user-defined observables and kinematical cuts, including a thorough assessment of the uncertainties related to scale and PDF variations. 
The Ntuple files are available upon request to the authors.

\section{Phenomenological results}

We present here selected results of interest for the LHC Run II, specifically NLO QCD predictions for the center-of-mass energy $\sqrt{s} = 13$ TeV.
To mimic as closely as possible the ATLAS and CMS detector acceptances, basic selection cuts are applied to the final states
\begin{equation}
\begin{array}{lclcl}
p_{T,\,\ell}>30 ~{\rm GeV} & \quad \quad \quad \quad
& p_{T,\,b}>40  ~{\rm GeV} &\quad \quad \quad \quad& p^{miss}_{T} >20
                                                     ~{\rm GeV} \\[0.2cm]
\Delta R_{\ell b} > 0.4 && \Delta R_{bb}>0.4 && \Delta R_{\ell \ell} > 0.4 \\[0.2cm]
|y_\ell|<2.5 &&  |y_b|<2.5 \,,  && 
\end{array}
\end{equation}
where $b$, $\ell$ denote respectively any $b$-jet and charged lepton.
Jets are defined according to the anti-$k_T$ clustering algorithm \cite{Cacciari:2008gp} with resolution parameter $R = 0.4$.
Our analysis requires exactly two b-jets, two charged leptons and missing $p_T$ in the final state. For the $t\bar{t}\gamma$ sample, we require additionally one isolated photon with $p_{T,\gamma} > 25$ GeV, $\vert y_\gamma \vert < 2.5$ and $\Delta R_{\ell \gamma}>0.4$.  According to the prescription of Ref. \cite{Frixione:1998jh}, the photon isolation condition reads 
\begin{equation}
\sum_{i} E_{T,\,i}  \, \Theta(R - R_{\gamma i})  \le E_{T,\,\gamma} \left(
\frac{1-\cos(R)}{1-\cos(R_{\gamma j})}
\right)\,,\\[0.2cm]
\end{equation}
where $R\le R_{\gamma j}=0.4$ and $i$ runs over all partons. 
For the renormalization and factorization scales two different prescriptions are considered, one being fixed and the other one being dynamic, \textit{i.e.} phase-space dependent. The fixed-scale choice is $\mu_R = \mu_F = m_t/2$ while the dynamic scale is defined to be $\mu_R = \mu_F = H_T/4$, where $H_T$ is the sum of the total missing $p_T$ plus the transverse momenta of the visible particles (\textit{i.e.} $b$-jets, charged leptons and, in the case of the $t\bar{t}\gamma$ sample, photons).

Reference values of the total cross sections obtained for the different scales considered in our analysis are reported in Table \ref{tab:1}. Results are shown for two different minimum requirements for the $p_T$ of the hard photon, namely 25 and 50 GeV. Looking at the total cross sections, both scale choices are in good shape and values agree well within the theoretical errors. We observe, however, that results based on the setup $\mu_0 = H_T/4$ feature systematically smaller scale uncertainties. In fact, as shown in Ref. \cite{Bevilacqua:2018woc} for the $t\bar{t}\gamma$ case, differential cross sections based on the $\mu_0 = H_T/4$ setup are characterized by moderate QCD corrections and reduced shape distortions induced by higher-order terms. This fact indicates that the proposed dynamical scale helps to capture effects from higher orders and accounts more effectively for the genuine multi-scale nature of the processes under consideration. We have calculated that the size of the internal uncertainties related to each of the three PDF sets considered lies in the range 1\--3\%, therefore smaller than the difference observed between various PDF sets which is at the level of 1\--4\%. In any case, the uncertainties stemming from scale variation represent the dominant source of theoretical systematics.

\begin{table}[t!]
\begin{center}
\begin{tabular}{c||c c||c}
&&&\\
 PDF set, $\mu_R=\mu_F=\mu_0$ & 
$\sigma^{\rm NLO}_{e^+\nu_e \mu^- \bar{\nu}_\mu
                    b\bar{b}}$ [fb]
& $\sigma^{\rm NLO}_{e^+\nu_e \mu^- \bar{\nu}_\mu b\bar{b}\gamma}$
  [fb]
  &$\sigma^{\rm NLO}_{e^+\nu_e \mu^- \bar{\nu}_\mu b\bar{b}\gamma}$
    [fb] \\[0.1cm]
&&$p_{T,\gamma} > 25$ GeV& $p_{T,\gamma} > 50$ GeV\\[0.1cm]
\hline \hline 
&&&\\
CT14, $\mu_0=m_t/2$ & $1629.4^{\,\,\,+18.4\,(1\%)}_{-144.7\,(9\%)}$  
& $7.436^{+0.074\,\,\,\,(1\%)}_{-1.034 \,(14\%)}$ 
& $3.081^{+0.050\,\,\,\,(2\%)}_{-0.514 \,(17\%)}$ \\[0.3cm]
CT14, $\mu_0=H_T/4$ & $1620.5^{\,\,\,+21.6\,\,(1\%)}_{-118.8\,\,(7\%)}$
&$7.496^{+0.099\,\,\,(1\%)}_{-0.457\,\,\,(6\%)}$ 
& $3.125^{+0.040\,\,\,(1\%)}_{-0.142\,\,\,(4\%)}$ 
\end{tabular}
\end{center}
\caption{\label{tab:1} \it  Total NLO cross sections for $pp \to e^+\nu_e \mu^- \bar{\nu}_\mu b\bar{b}$ ("$t\bar{t}$") and $pp \to e^+\nu_e \mu^- \bar{\nu}_\mu b\bar{b} \gamma$ ("$t\bar{t}\gamma$"), for different scale choices \cite{Bevilacqua:2018dny}. Also included are theoretical errors as obtained from scale variation. In the case of $t\bar{t}\gamma$, results for two different values of the $p_{T,\gamma}$ cut are reported.}
\end{table}

\begin{figure}
\begin{center}
\includegraphics[width=.42\textwidth]{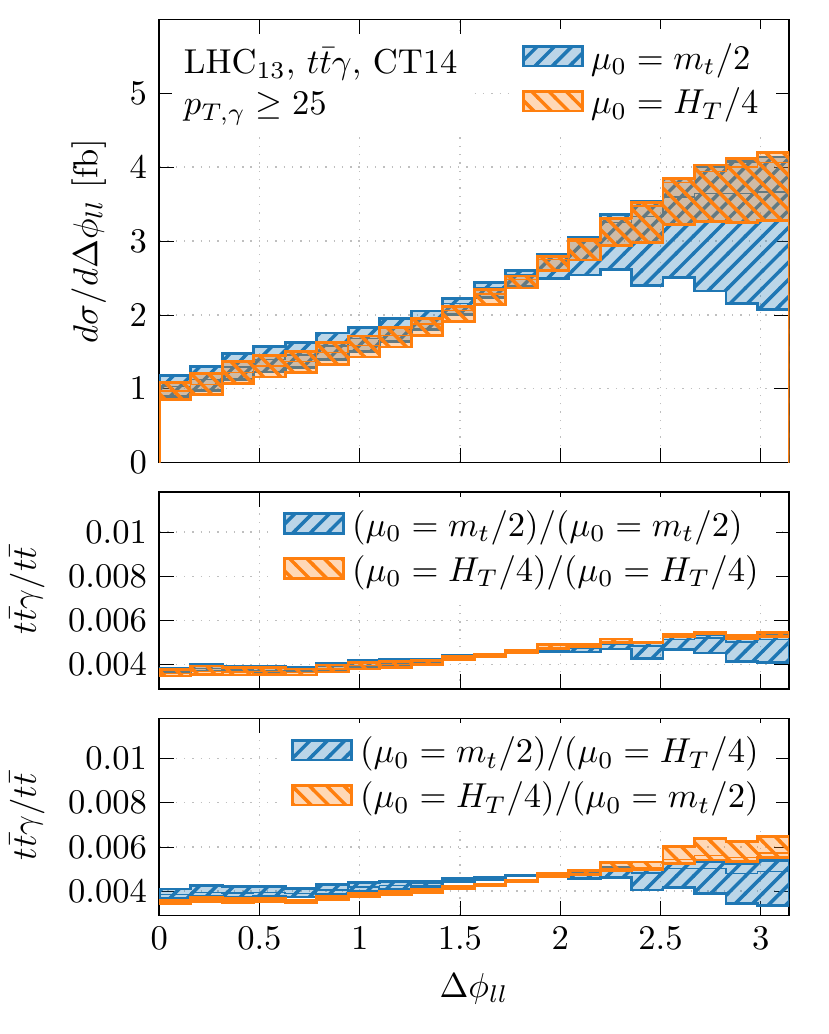} 
\includegraphics[width=.42\textwidth]{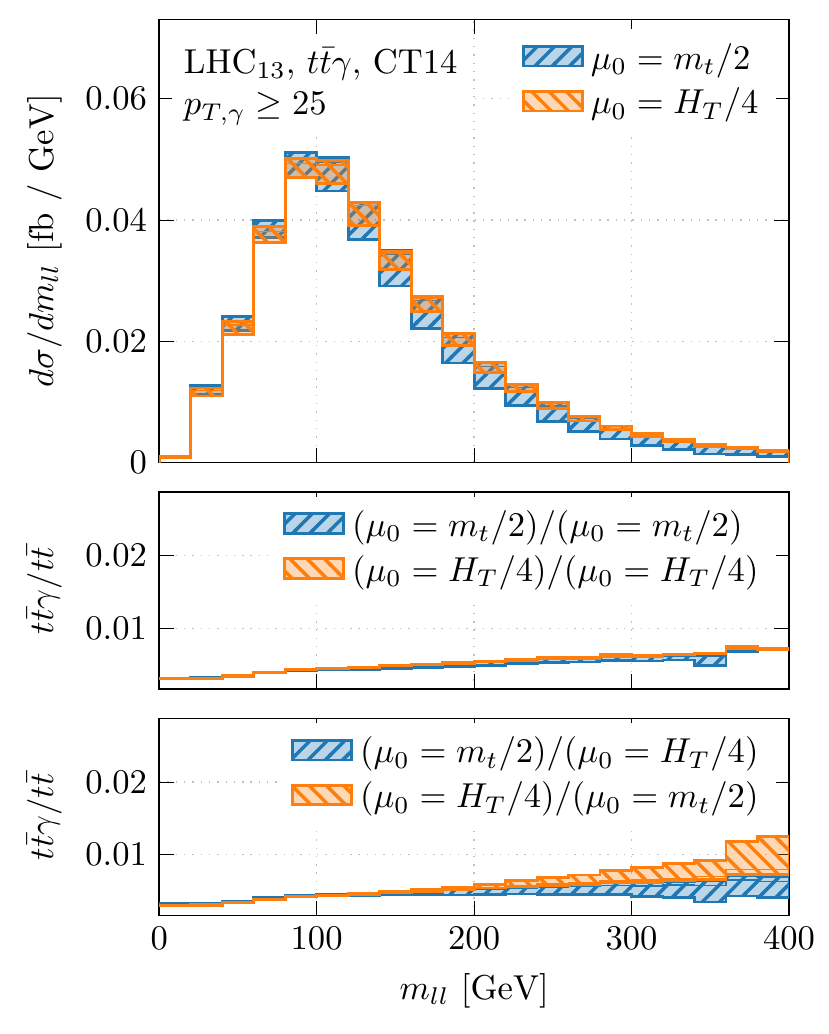} 
\caption{Impact of scale uncertainties at NLO: comparison between absolute $t\bar{t}\gamma$ predictions and the $t\bar{t}\gamma/t\bar{t}$ ratio \cite{Bevilacqua:2018dny}. \textit{Upper panels}: absolute $t\bar{t}\gamma$ predictions.  \textit{Middle panels}: predictions for the $t\bar{t}\gamma/t\bar{t}$ ratio using same scale for numerator and denominator. \textit{Lower panels}: predictions for the $t\bar{t}\gamma/t\bar{t}$ ratio using different scales for numerator and denominator.
The observables shown are the azimuthal angle between the two charged leptons (on the left) and the invariant mass of the the two-lepton system (on the right).
}
\label{fig:ttA_abs_vs_ratio}
\end{center}
\end{figure}

In Ref. \cite{Bevilacqua:2018dny} we have performed a detailed analysis of correlations between the $t\bar{t}$ and $t\bar{t}\gamma$ processes. Specifically, we looked at the shapes of NLO distributions for several observables, both dimensionless and dimensionful, to outline possible kinematical differences. We concluded that $t\bar{t}$ and $t\bar{t}\gamma$ production show strong similarities in both jet and leptonic activity and can be considered correlated as far as these observables are concerned.
Invoking correlations, one may restrict the set of combinations of scale variations used to estimate the uncertainty of the ratio, 
$\mathcal{R} = \frac{\sigma^{NLO}_{t\bar{t}\gamma}(\mu_1)}{\sigma^{NLO}_{t\bar{t}}(\mu_2)}$,
so that $(\mu_1,\mu_2)$ are either increased or decreased at the same time. 
Following this prescription the uncertainties estimated from scale variation could be dramatically reduced. 
Figure \ref{fig:ttA_abs_vs_ratio} shows a comparison between absolute $t\bar{t}\gamma$ predictions and the $t\bar{t}\gamma$\,/\,$t\bar{t}$ ratio for two representative observables, namely the azimuthal angle between the two charged leptons ($\Delta \phi_{ll}$) and the invariant mass of the lepton pair ($m_{ll}$). 
Looking at the first observable, the absolute $t\bar{t}\gamma$ predictions based on $\mu_0 = H_T/4$ are characterized by uncertainties of the order of 15\% in the region $\Delta \phi_{ll} \approx 3$.
The cross section ratio features a dramatic reduction down to 3\% in the same region when correlations are invoked. The results for the $m_{ll}$ observable are on the same footing, as one observes scale uncertainties of a few percents in the case of the correlated ratio. For comparison, in the bottom panels of Figure \ref{fig:ttA_abs_vs_ratio} we show an example of assigning different scales in the numerator and the denominator of the cross section ratio. The uncertainties obtained in this way are much larger and reach the order of 15\% \-- 40\% depending on the scale assignment.

\section{Conclusions}

The purpose of this work was to investigate whether more precise predictions for $t\bar{t}\gamma$ production in the dilepton decay channel at the LHC can be obtained using state-of-the-art theoretical accuracy, given that full NNLO results  are still out of reach. To this end we have studied the cross section ratio $\mathcal{R} = \sigma^{NLO}_{t\bar{t}\gamma}/\sigma^{NLO}_{t\bar{t}}$ and performed a thourough assessment of the theoretical uncertainties associated to scale and PDF dependence. Our best predictions for this observable can be summarized as follows:
\begin{eqnarray}
{\cal R}(\mu_0=H_T/4, \, p_{T,\gamma} > 25 \,{\rm GeV} ) &= (4.62 \, \pm \, 0.06 \, [\mbox{scale}] \, \pm \, 0.04 \, [\mbox{PDF}] ) \cdot 10^{-3} \,, \nonumber \\ [0.1cm] 
{\cal R}(\mu_0=H_T/4, \, p_{T,\gamma} > 50 \,{\rm GeV} ) &= (1.93 \, \pm \, 0.06 \, [\mbox{scale}] \, \pm \, 0.04 \, [\mbox{PDF}] ) \cdot 10^{-3} \,. \nonumber
\end{eqnarray}
With uncertainties that can be pinned down to a few percents, a very precise determination of this observable is possible. We conclude that the cross section ratio has an interesting potential to shed light on new physics that can reveal itself only when sufficiently precise theoretical predictions are available.



\begin{thebibliography}{99}

\bibitem{Czakon:2013goa}
  M.~Czakon, P.~Fiedler and A.~Mitov,
  Phys.\ Rev.\ Lett.\  {\bf 110} (2013) 252004
  [arXiv:1303.6254 [hep-ph]].

\bibitem{Czakon:2017wor}
  M.~Czakon, D.~Heymes, A.~Mitov, D.~Pagani, I.~Tsinikos and M.~Zaro,
  JHEP {\bf 1710} (2017) 186
  [arXiv:1705.04105 [hep-ph]].

\bibitem{Behring:2019iiv}
  A.~Behring, M.~Czakon, A.~Mitov, A.~S.~Papanastasiou and R.~Poncelet,
  arXiv:1901.05407 [hep-ph].

\bibitem{Sirunyan:2018hoz}
  A.~M.~Sirunyan {\it et al.} [CMS Collaboration],
  Phys.\ Rev.\ Lett.\  {\bf 120} (2018) no.23,  231801
  [arXiv:1804.02610 [hep-ex]].

\bibitem{Sirunyan:2017uzs}
  A.~M.~Sirunyan {\it et al.} [CMS Collaboration],
  JHEP {\bf 1808} (2018) 011
  [arXiv:1711.02547 [hep-ex]].

\bibitem{Aaboud:2018hip}
  M.~Aaboud {\it et al.} [ATLAS Collaboration],
  Eur.\ Phys.\ J.\ C {\bf 79} (2019) no.5,  382
  [arXiv:1812.01697 [hep-ex]].

\bibitem{PengFei:2009ph}
  P.~F.~Duan \textit{et al.},
  Phys.\ Rev.\ D {\bf 80} (2009) 014022
  [arXiv:0907.1324 [hep-ph]].

\bibitem{PengFei:2011qg}
  P.~F.~Duan \textit{et al.},
  Chin.\ Phys.\ Lett.\  {\bf 28} (2011) 111401
  [arXiv:1110.2315 [hep-ph]].

\bibitem{Duan:2016qlc}
  P.~F.~Duan \textit{et al.},
  Phys.\ Lett.\ B {\bf 766} (2017) 102
  [arXiv:1612.00248 [hep-ph]].

\bibitem{Maltoni:2015ena}
  F.~Maltoni, D.~Pagani and I.~Tsinikos,
  JHEP {\bf 1602} (2016) 113
  [arXiv:1507.05640 [hep-ph]].

\bibitem{Melnikov:2011ta}
  K.~Melnikov, M.~Schulze and A.~Scharf,
  Phys.\ Rev.\ D {\bf 83} (2011) 074013
  [arXiv:1102.1967 [hep-ph]].
  
 \bibitem{Kardos:2014zba}
  A.~Kardos and Z.~Trocsanyi,
  JHEP {\bf 1505} (2015) 090
  [arXiv:1406.2324 [hep-ph]].

\bibitem{Baur:2004uw}
  U.~Baur, A.~Juste, L.~H.~Orr and D.~Rainwater,
  Phys.\ Rev.\ D {\bf 71} (2005) 054013
  [hep-ph/0412021].
  
\bibitem{Bylund:2016phk}
  O.~Bessidskaia Bylund, F.~Maltoni, I.~Tsinikos, E.~Vryonidou and C.~Zhang,
  JHEP {\bf 1605} (2016) 052
  [arXiv:1601.08193 [hep-ph]].

\bibitem{Schulze:2016qas}
  M.~Schulze and Y.~Soreq,
  Eur.\ Phys.\ J.\ C {\bf 76} (2016) no.8,  466
  [arXiv:1603.08911 [hep-ph]].

\bibitem{Bevilacqua:2018woc}
  G.~Bevilacqua, H.~B.~Hartanto, M.~Kraus, T.~Weber and M.~Worek,
  JHEP {\bf 1810} (2018) 158
  [arXiv:1803.09916 [hep-ph]].

\bibitem{Bevilacqua:2018dny}
  G.~Bevilacqua, H.~B.~Hartanto, M.~Kraus, T.~Weber and M.~Worek,
  JHEP {\bf 1901} (2019) 188
  [arXiv:1809.08562 [hep-ph]].

\bibitem{Denner:2010jp}
  A.~Denner, S.~Dittmaier, S.~Kallweit and S.~Pozzorini,
  Phys.\ Rev.\ Lett.\  {\bf 106} (2011) 052001
  [arXiv:1012.3975 [hep-ph]].

\bibitem{Bevilacqua:2010qb}
  G.~Bevilacqua, M.~Czakon, A.~van Hameren, C.~G.~Papadopoulos and M.~Worek,
  JHEP {\bf 1102} (2011) 083
  [arXiv:1012.4230 [hep-ph]].

\bibitem{Denner:2012yc}
  A.~Denner, S.~Dittmaier, S.~Kallweit and S.~Pozzorini,
  JHEP {\bf 1210} (2012) 110
  [arXiv:1207.5018 [hep-ph]].

\bibitem{Mangano:2012mh}
  M.~L.~Mangano and J.~Rojo,
  JHEP {\bf 1208} (2012) 010
  [arXiv:1206.3557 [hep-ph]].

\bibitem{Bevilacqua:2014qfa}
  G.~Bevilacqua and M.~Worek,
  JHEP {\bf 1407} (2014) 135
  [arXiv:1403.2046 [hep-ph]].

\bibitem{Bevilacqua:2011xh}
  G.~Bevilacqua, M.~Czakon, M.~V.~Garzelli, A.~van Hameren, A.~Kardos, C.~G.~Papadopoulos, R.~Pittau and M.~Worek,
  Comput.\ Phys.\ Commun.\  {\bf 184} (2013) 986
  [arXiv:1110.1499 [hep-ph]].

\bibitem{vanHameren:2009dr}
  A.~van Hameren, C.~G.~Papadopoulos and R.~Pittau,
  JHEP {\bf 0909} (2009) 106
  [arXiv:0903.4665 [hep-ph]].

\bibitem{Czakon:2009ss}
  M.~Czakon, C.~G.~Papadopoulos and M.~Worek,
  JHEP {\bf 0908} (2009) 085
  [arXiv:0905.0883 [hep-ph]].
  
\bibitem{Bevilacqua:2013iha}
  G.~Bevilacqua, M.~Czakon, M.~Kubocz and M.~Worek,
  JHEP {\bf 1310} (2013) 204
  [arXiv:1308.5605 [hep-ph]].

\bibitem{Bern:2013zja}
  Z.~Bern \textit{et al.},
  Comput.\ Phys.\ Commun.\  {\bf 185} (2014) 1443
  [arXiv:1310.7439 [hep-ph]].
  
\bibitem{Butterworth:2015oua}
  J.~Butterworth {\it et al.},
  J.\ Phys.\ G {\bf 43} (2016) 023001
  [arXiv:1510.03865 [hep-ph]].

\bibitem{Dulat:2015mca}
  S.~Dulat {\it et al.},
  Phys.\ Rev.\ D {\bf 93} (2016) no.3,  033006
  [arXiv:1506.07443 [hep-ph]].

\bibitem{Harland-Lang:2014zoa}
  L.~A.~Harland-Lang \textit{et al.},
  Eur.\ Phys.\ J.\ C {\bf 75} (2015) no.5,  204
  [arXiv:1412.3989 [hep-ph]].

\bibitem{Ball:2014uwa}
  R.~D.~Ball {\it et al.} [NNPDF Collaboration],
  JHEP {\bf 1504} (2015) 040
  [arXiv:1410.8849 [hep-ph]].

\bibitem{Cacciari:2008gp}
  M.~Cacciari, G.~P.~Salam and G.~Soyez,
  JHEP {\bf 0804} (2008) 063
  [arXiv:0802.1189 [hep-ph]].

\bibitem{Frixione:1998jh}
  S.~Frixione,
  Phys.\ Lett.\ B {\bf 429} (1998) 369
  [hep-ph/9801442].

\end{thebibliography}
\end{document}